\documentclass[runningheads]{llncs}
\usepackage{graphicx}

\begin{document}

\title{DN-ResNet: Efficient Deep Residual Network for Image Denoising} 
\titlerunning{DN-ResNet: Efficient Deep Residual Network for Image Denoising} 


\author{Haoyu Ren, Mostafa El-khamy, Jungwon Lee}


\institute{SOC R\&D, Samsung Semiconductor Inc.\\
	4921 Directors Pl, San Diego, CA 92121, USA \\
	\email{ \{haoyu.ren,mostafa.e,jungwon2.lee\}@samsung.com}
}
%




\maketitle

\begin{abstract}
A deep learning approach to blind denoising of images without complete knowledge of the noise statistics is considered. We 
propose DN-ResNet, which is a deep convolutional neural network (CNN) consisting of several residual blocks (ResBlocks). With 
cascade training, DN-ResNet is more accurate and more computationally efficient than the state of art denoising networks. An 
edge-aware loss function is further utilized in training DN-ResNet, so that the denoising results have better perceptive 
quality compared to conventional loss function. Next, we introduce the depthwise separable DN-ResNet (DS-DN-ResNet) utilizing 
the proposed Depthwise Seperable ResBlock (DS-ResBlock) instead of standard ResBlock, which has much less computational cost. DS-DN-ResNet is incrementally evolved by replacing the ResBlocks in DN-ResNet by DS-ResBlocks stage by stage. As 
a result, high accuracy and good computational efficiency are achieved concurrently. Whereas previous state of art deep 
learning methods focused on denoising either Gaussian or Poisson corrupted images, we consider denoising images having the more 
practical Poisson with additive Gaussian noise as well. The results show that DN-ResNets are more efficient, robust, and 
perform better denoising than current state of art deep learning methods, as well as the popular variants of the BM3D 
algorithm, in cases of blind and non-blind denoising of images corrupted with Poisson, Gaussian or Poisson-Gaussian noise. Our 
network also works well for other image enhancement task such as compressed image restoration.

\end{abstract}

\section{Introduction}
\noindent{}Denoising is an active topic in image processing since it is a key step in many practical applications, such as image and video capturing. It aims to generate a clean image ${X}$ from a given noisy image ${Y}$ which follows an image degradation model ${Y}=D({X})$. For the widely used additive Gaussian noise (AWGN) model, the $i${th} observed pixel is 

\begin{equation}
y_i = D(x_i) = x_i + n_i
\end{equation}

\noindent{}where $n_i \sim  \mathcal{N}(0, \sigma^2)$ is i.i.d Gaussian noise with zero mean and variance $\sigma^2$. AWGN has been used to model the signal-independent thermal noise and other system imperfections. Degradation due to low light shot noise is signal dependent and has often been modeled using Poisson noise

\begin{equation}
y_i=D(x_i) = p_i, \quad p_i \sim \mathcal{P}(x_i)
\end{equation}

\noindent{}where $\mathcal{P}(x_i)$ is a Poisson random variable with mean $x_i$. However, this noise approaches a Gaussian distribution for average light conditions as $\mathcal{P}(\lambda) \approx \mathcal{N}(\lambda, \lambda)$, for large enough $\lambda$. Hence, the noise due to capturing by an imaging device is better modeled as a Poisson noise with AWGN, referred to as Poisson-Gaussian noise, such that 

\begin{equation}
y_i = D(x_i) = \alpha p_i + n_i, \quad \alpha > 0
\end{equation}

\noindent{}which has been verified by experimental results \cite{foi2008practical}.
%
%

Recently, the state of art denoising accuracy is achieved by deep neural networks \cite{zhang2017beyond}\cite{tai2017memnet}, which construct a mapping between the noisy image and clean image. Unforunately, most of existing denoising networks cannot be executed in real-time due to their large network size. In addition, it is relatively difficult to set the hyperparameters when learning a very deep network, such as the weight initialization, the learning rate, and the weight decay rate. With inappropriate parameters, the training might fall into local minimum or not converge at all.

In this paper, we propose a Denoising Residual Network (DN-ResNet) which is more efficient and accurate than prior art. 
DN-ResNet consists of residual blocks (ResBlock) which are gradually inserted into the network stage by stage during the training. This training strategy not only allows the resulting DN-ResNet to converge faster, but also allows it to be more computationally efficient than prior art denoising networks. Even better perceptive quality have been observed by using the proposed edge-aware loss function instead of the conventional mean square error (MSE). In addition, we introduce the depthwise separable ResBlock (DS-ResBlock) into DN-ResNet to construct the depthwise separable ResNet (DS-DN-ResNet). DS-DN-ResNet is generated by the proposed incremental evolution from DN-ResNet, where the ResBlocks in DN-ResNet are replaced by DS-ResBlocks stage by stage. As a result, we may obtain a 2.5 times complexity reduction for DN-ResNet, with less than 0.1 dB PSNR loss. To our knowledge, DN-ResNet is the first unified deep CNN trained for the problem of blind denoising of images corrupted by multiple type of noises. By cascading only 5 ResBlocks, DN-ResNet and DS-DN-ResNet achieve the state of art performance on all three denoising problems, Gaussian, Poisson, and Poisson-Gaussian, for both cases of non-blind denoising (known noise level for noisy input) and blind denoising (unknown noise level for noisy input). The speed is also much faster than prior art denoising networks. Moreover, we show that DN-ResNet works well for compressed image restoration. This implies that DN-ResNet can be generalized to other applications.


%
%
As summary, our contributions are three folds:

\noindent{}1. We show that ResNet is effective for image denoising, and using edge-aware loss function significantly improves the perceptive quality. The resulting DN-ResNet achieves the state of art accuracy, and is 4 times less complicate than existing networks;

\noindent{}2. We introduce the depthwise separable ResBlock (DS-ResBlock) to construct DS-DN-ResNet. The incrementally evolved DS-DN-ResNet is 2.5 times faster than DN-ResNet, without significant accuracy loss;

\noindent{}3. We show that the proposed DN-ResNet works well for all types of noises, even without knowing the noise level. It can be generalized to other image enhancement task such as compressed image restoration;

\section{Related work}
\subsection{Image denoising}
\noindent{}During the past years, numerous approaches have been exploited for modeling image priors for denoising, such as nonlocal self-similarity (NSS) \cite{gu2014weighted} and sparse coding \cite{dong2013nonlocally}. The block matching with 3D collaborative filtering (BM3D) \cite{dabov2007image} and its variants such as iterative BM3D with variance stabilizing transforms (I+VST+BM3D) \cite{azzari2016variance} and generalized Anscombe variance stabilizing transform with BM3D (GAT-BM3D) \cite{makitalo2013optimal} are widely used. These methods generally involve a complex optimization problem in the testing stage, which makes the denoising process time-consuming. To improve the efficiency, learning-based  methods are proposed to get rid of the iterative optimization procedure, such as the trainable nonlinear reaction diffusion (TNRD) \cite{chen2017trainable}, and Gaussian conditional random field \cite{schmidt2016cascades} for non-blind image deblurring. Unforunately, the accuracy of these methods is still limited due to the use of specific image prior. It is also difficult to set the handcrafted parameters during the stage-wise learning. 

Recently, deep neural networks have been deployed for image denoising due to their significant improvement of the accuracy \cite{burger2012image}. Zhang et al. \cite{zhang2017beyond} constructed a 20-layer feed-forward denoising convolutional neural networks with residual learning for Gaussian denoising. Remez et al. trained 20-layer CNNs for each object category respectively and showed good performance for either Gaussian denoising \cite{remez2017deep} or Poisson denoising \cite{remez2017deep2}. Zhang et al. \cite{zhang2017ffdnet} proposed FFDNet adopting orthogonal regularization to enhance the generalization ability of Gaussian denoising. Tai et al. designed MemNet \cite{tai2017memnet}, where the feature map concatenations and skip connections are utilized to construct a network for image super resolution, Gaussain denoising, and JPEG deblocking. $1\times1$ convolutions are adopted to integrate the long-term memorization, which shows significant accuracy improvements. Most of the existing networks are designed for single type of noise only. Due to the high computational cost, they can not be executed in real-time. In contrast, our DN-ResNet is far more efficient. The same network architecture can be utilized for Gaussian, Poisson, and Poisson-Gaussian noise, as well as other image enhancement tasks.

\subsection{Deep learing based compressed image restoration}
\noindent{}Compressed image restoration aims to reduce the artifacts of decoded compressed images, so that the images can be stored or transmitted at low bit rates. Most of existing work design an end-to-end network including both the encoding (compression) and decoding procedure. Toderici et al. \cite{toderici2017full} presented a set of full-resolution lossy image compression methods using recurrent neural network based encoder and decoder with entropy coding. Theis et al. \cite{theis2017lossy} constructed the compression network by deep autoencoders with a sub-pixel structure. In these work, although a low bit rate can be achieved, both of the encoding and decoding procedure are replaced by deep neural networks. As a result, it is difficult to integrate them into real system, where efficient image compression algorithms such as JPEG are implemented. In this paper, we consider the compressed image restoration as a `denoising' problem, where the noise comes from image compression algorithms. DN-ResNet is trained to refine the quality of decoded compressed image. Since our network can be considered as a post-processing step, it can be applied to any existing image compression algorithms.

\section{Denoising Residual Network}
\subsection{DN-ResNet}
\noindent{}We aim to train a deep convolution neural network for image denoising. The network takes a noisy image ${Y}$ as input and predicts a clean image ${X}$ as its output. Given a training set $\{{X}_i, Y_i\}, i = 1,\ldots,N$ with $N$ samples, our goal is to learn a model $S$ that predicts the  clean image $\hat{X_i} = S({Y}_i)$. 

ResNet \cite{he2016deep} has demonstrated considerable performance in computer vision applications such as image classification. The basic element of our proposed denoising residual network (DN-ResNet) is a simplified ResBlock, as shown in Fig. 1(b). Different from the standard ResBlock in Fig. 1(a), we remove the batch normalization layers and the ReLU layer after the addition, because removing these layers will not harm the performance of feature-map based ResNet \cite{lim2017enhanced}. 

\begin{figure}
\centering
\includegraphics[height=4.5cm]{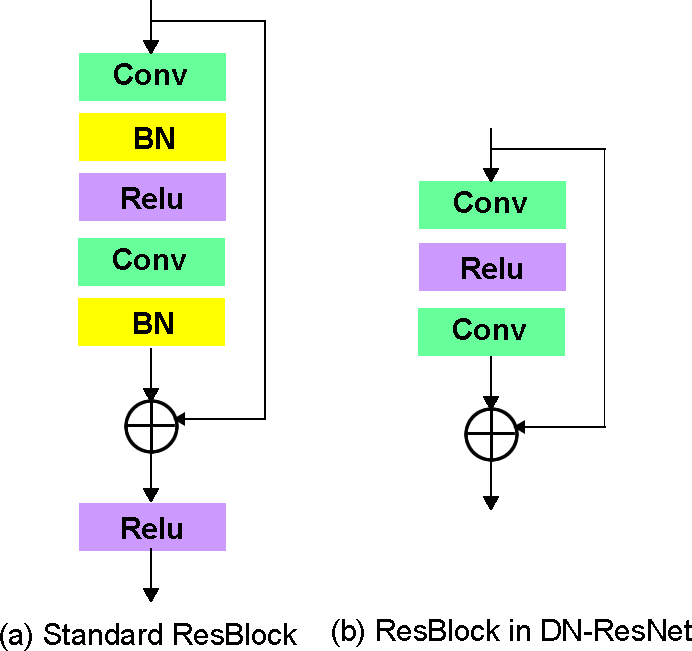}
\caption{ResBlocks in DN-ResNet. (a) Standard ResBlock (b) ResBlock in DN-ResNet. }
\label{fig:example}
\end{figure}

We construct our DN-ResNet by concatenating the ResBlocks in Fig. 1(b). We observed that as the network goes deeper, the training and the hyper-parameter tuning become more difficult. To solve this problem, we follow the cascade training \cite{ren2017ct}, which separates the whole training into stages and proceeds one by one. The training of DN-ResNet starts from a simple 3-layer CNN model. The first layer consists of 64 $9\times9$ filters. The second layer consist of 32 $5\times5$ filters. There is only one $5\times5$ filter in the last layer. All convolutions have stride one, and all the weights are randomly initialized from a Gaussian distribution with $\sigma = 0.001$. After the 3-layer CNN is trained, we start cascading the ResBlocks stage by stage, as shown in Fig. 2. When the training of current stage is finished, e.g., the training loss of current stage is 3\% lower than previous stage, the training will proceed to next stage, and the network is cascaded to a deeper network. In each stage, one new ResBlock is inserted. So the training starts from 3 layers, and proceeds to 5 layers, 7 layers, etc.. Each convolutional layer in the ResBlock consists of 32 $3\times3$ filters. It ensures a smaller network when going deeper. The new layers are inserted just before the last $5\times5$ layer. The weights of pre-existing layers are inherited from the previous stage, and the weights of the new  ResBlocks are randomly initialized (Gaussian with $\sigma = 0.001$).  Hence, only a few weights of DN-ResNet are randomly initialized at each stage, so the convergence will be relatively easy. We find that using a fixed learning rate 0.0001 for all layers without any decay is feasible. 

Since new convolutional layers will reduce the size of the feature map, we zero pad 2 pixels in each new $3\times{}3$ layer. As a result, all the stages in cascade training have the same size as the output, so that the training samples could be shared.
When cascading 5 ResBlocks, the resulting DN-ResNet will have $5\times2+3=13$ convolutional layers. Our experiments show that such DN-ResNet-13 has already achieved the state of art accuracy on all type of noises.

\begin{figure}
\centering
\includegraphics[height=4.5cm]{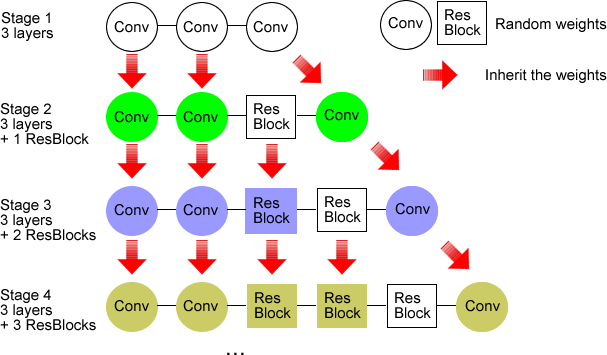}
\caption{Cascade training of DN-ResNet. Circle denotes standard convolutional layer, rectangle denotes ResBlock.}
\label{fig:example}
\end{figure}

\subsection{Depthwise separable DN-ResNet}
\noindent{}In this section, we propose depthwise separable DN-ResNet (DS-DN-ResNet) to further reduce the network size of DN-ResNet, as well as the computational cost. In the classification network MobileNet \cite{howard2017mobilenets}\cite{sandler2018mobilenetv2}, the standard convolutional layer is factorized into a depthwise convolution and a $1\times1$ pointwise convolution, which achieves significant efficiency gain. As shown in Fig. 3, the standard convolution with $M$ input channels and $N$ ${K}\times{K}$ filters is replaced by a depthwise convolutional layer with $M$ $K\times{}K$ filters, and a pointwise convolutional layer with $N$  $1\times1$ convolutional filters and $M$ input channels. Assume the input feature map size is $W\times{}H$, the number of the multiplications are reduced from $M\times{}K\times{}K\times{}N\times{}W\times{}H$ to $M\times{}K\times{}K\times{}W\times{}H+M\times{}N\times{}W\times{}H$. 


\begin{figure}
\centering
\includegraphics[height=4.5cm]{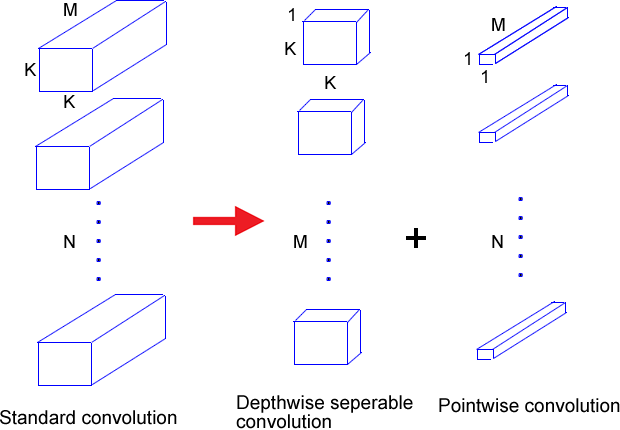}
\caption{Depthwise separable convolution. The standard convolution (left) is replaced by depthwise convolution (middle) and pointwise convolution (right).}
\label{fig:example}
\end{figure}

Inspired by this idea, we propose the depthwise separable ResBlock (DS-ResBlock), as shown in Fig. 4. In DS-ResBlock, the standard convolutional layers in ResBlock are replaced by depthwise separable convolutional layers and pointwise convolutional layers.  Relu activation is added for all the convolutional layers in DS-ResBlock. Assume the size of input feature map is $640\times480$, the number of the multiplications  in the ResBlock in Fig. 4(a) is $640\times480\times3\times3\times32\times32\times2=5.6\times10^9$. In the corresponding DS-ResBlock in Fig. 4(b), the number of multiplications is $640\times480\times3\times3\times32 + 640\times480\times32\times32=9\times10^8$. The computational cost is reduced 6 times.

\begin{figure}
\centering
\includegraphics[height=4.5cm]{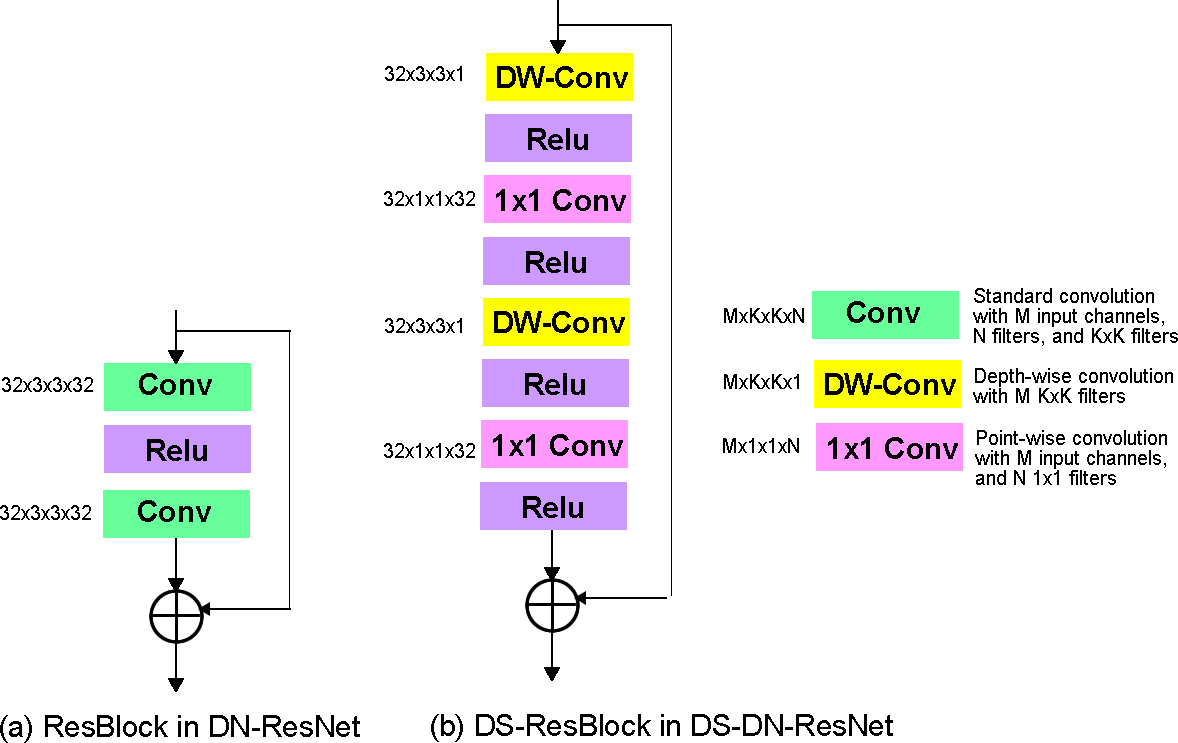}
\caption{Comparison between the ResBlock in DN-ResNet and the DS-ResBlock in DS-DN-ResNet.}
\label{fig:example}
\end{figure}

To train DS-DN-ResNet, one intuitive way is to apply cascade training from scratch. Since we have already trained the DN-ResNet, we describe another way to obtain DS-DN-ResNet to save training time, which is called `incremental evolution'. To obtain a DS-DN-ResNet from existing DN-ResNet, a feasible way is to replace all ResBlocks by DS-ResBlocks, and fine-tune the whole network. If this procedure is done in a one-shot way, the fine-tuning will not converge well (see Table 3 for details). In the incremental evolution, the ResBlocks are replaced stage by stage. In each stage, only one-ResBlock is replaced by DS-ResBlock, and followed by a fine-tuning, as shown in Fig. 5. Similar to cascade training, the weights in the new DS-ResBlock are randomly initialized, and the weights in all other layers are inherited. The replacement starts from the tail side to ensure a smaller influence to the whole network. In the implementation, we first train 13-layer DN-ResNet, and then evolve it to DS-DN-ResNet. The learning rate is the same as cascade training. The fine-tuning will last 10 epochs for each evolution stage.

After incremental evolution, there are still three standard convolutional layers (1st, 2nd, and the last one) in DS-DN-ResNet. The overall complexity of DS-DN-ResNet is about 2.5 times less compared to DN-ResNet, without significant accuarcy loss ($<0.1$ dB PSNR, see Table 3 for details). We do not replace the 1st and 2nd standard convolutional layers by depthwise version because it will decrease the accuracy a lot ($>0.3$ dB PSNR).  


\begin{figure}
\centering
\includegraphics[height=6cm]{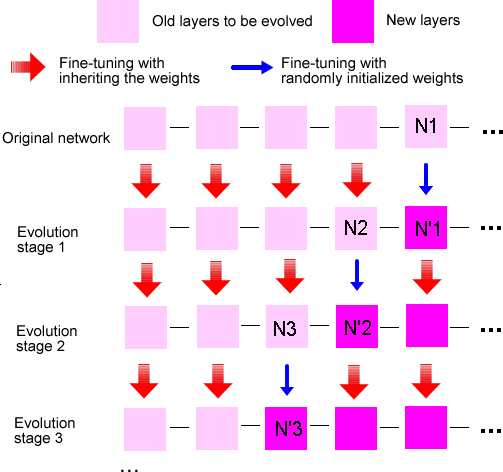}
\caption{Incremental evolution from DN-ResNet to DS-DN-ResNet.}
\label{fig:example}
\end{figure}

\subsection{Edge-aware loss function}
\noindent{}Most of existing denoising networks aim to minimize the Mean Square Error (MSE) $\frac{1}{N}\sum_{i=1}^N||X_i-\hat{X}_i||^2$ over the training set. In this paper, we propose an edge-aware loss function, where the pixels in the edges are granted higher weights compared to non-edge pixels

\begin{equation}
loss = \frac{1}{N}\sum_{i=1}^N||X_i-\hat{X}_i||^2 + w\times{}\frac{1}{N}\sum_{i=1}^{N}||X_i{}M_i-\hat{X}_i{}M_i||^2.
\end{equation}

\noindent{}In Eq. (4), $X_i$ is the ground truth of $i$th clean image, $\hat{X_i}$ is the $i$th denoised image, $M$ is an edge map, $N$ is the number of images, and $w$ is a constant to control the trade-off between edge and non-edge pixels. 

There are two advantages of applying such edge-aware loss function. Firstly, one of the major challenge in image denoising is that the edges are difficult to be retrieved from a noisy image, especially when the noise level is high. Adding a  corresponding constraint in the loss function is reasonable. Secondly, the high-frequency information such as edge is very sensitive in human vision. Increasing the denoising accuracy of edge pixels will contribute to the perceptive quality.  

We try two ways to construct $M$, including (a) gradient magnitude from Sobel filter, and (b) binary edge mask by thresholding (a). In the experiments, we show that using such edge-aware loss function can grant us better perceptive quality. The SSIM (structual similarity measure) is significantly improved.


\begin{table*}[h]
\caption{PSNR (dB) evaluation of DN-ResNets with different layers for different denoising problens on PASCAL VOC 2010 dataset. Noise level is known. Bold fold indicates the best. Conventional MSE loss function is utilized for all models. }
\scriptsize
\renewcommand{\arraystretch}{1.0}
\centering
\begin{tabular}{|c|c|c|c|c|c|c|c|c|}
\hline
{DN-ResNet} & {sigma/peak} & {3-layer} & 5-layer &{7-layer} &{9-layer} &{11-layer} & {13-layer} & {13-layer-os}  \\ \hline
Parameters & - & 57,184 & 75,616 & 94,048 & 112,480 & 130,912 & 149,344 & 149,344 \\ \hline
& 10 & 34.43 & 34.56 & 34.71& 34.80 & 34.93  & \textbf{34.99} & 34.70  \\
Gaussian  & 25 & 29.86 & 30.03 & 30.10 & 30.30 & 30.44 & \textbf{30.52} & 30.27  \\ 
& 50 & 26.86 & 27.05 & 27.22 & 27.29 & 27.38 & \textbf{27.50} & 27.14\\ 
& 75 & 25.24 & 25.43 & 25.55 & 25.63 & 25.81 & \textbf{25.89} & 25.61 \\ \hline
& 1 &  22.51 & 22.66 & 22.74 & 22.88 & 22.95 & \textbf{23.06} & 22.80  \\
Poisson & 2 & 23.66 & 23.74 & 23.92 & 24.05 & 24.14 & \textbf{24.23} & 23.96  \\ 
& 4 &  24.67 & 24.80 & 24.91 & 25.14 & 25.27 & \textbf{25.39} & 25.01  \\
& 8 &  26.01 & 26.24 & 26.35 & 26.55 & 26.64 & \textbf{26.77} & 26.49  \\ \hline
& 0.1/1 & 22.11 & 22.27 & 22.36 & 22.50 & 22.65 & \textbf{22.73} & 22.30 \\
& 0.2/2 & 22.99 & 23.14 & 23.22 & 23.40 & 23.59 & \textbf{23.75} & 23.44 \\
& 0.5/5 & 24.54 & 24.61 & 24.77 & 24.90 & 25.00 & \textbf{25.10} & 24.78 \\
Poisson-Gaussian & 1/10 & 25.61 & 25.69 & 25.77 & 25.91 & 25.99 & \textbf{26.14} & 25.67 \\
& 2/20 &  26.59 & 26.71 & 26.89 & 26.99 & 27.14 & \textbf{27.29} & 26.88 \\
& 3/30 &  27.10 & 27.22 & 27.37 & 27.50 & 27.61 & \textbf{27.77} & 27.41 \\
& 6/60 & 27.87 & 27.98 & 28.16 & 28.32 & 28.48 & \textbf{28.59} & 28.11 \\
& 12/120 & 28.19 & 28.30 & 28.44 & 28.58 & 28.72 & \textbf{28.88} & 28.50 \\ \hline
\end{tabular}
\end{table*}

\begin{table*}[ht]
\caption{PSNR (dB)/SSIM evaluation of 13-layer DN-ResNets with different loss functions and blind/non-blind denoising. Bold fold indicates the best SSIM. `edge' means the network is trained by edge-aware loss function. In `edge-a', the edge map in the edge-aware loss function is the gradient magnitude from Sobel filter. In `edge-b', the edge map is a binary mask by thresholding the Sobel map with 150.  We have tried different thresholds to obtain the binary edge mask and find that 150 is the best one.}
\scriptsize
\renewcommand{\arraystretch}{1.0}
\centering
\begin{tabular}{|c|c|c|c|c|c|}
\hline
{DN-ResNet} & sigma/peak & {non-blind} & {blind} & {blind+`edge-a'} & blind+`edge-b' \\ \hline
Parameters & - & 149,344 & 149,344 & 149,344 & 149,344  \\ \hline
& 10 & {34.99/0.9224} & 34.88/0.9217 & 34.88/\textbf{0.9271} & 34.85/0.9266  \\
Gaussian  & 25 &  {30.52/0.8383} & 30.47/0.8369 & 30.45/\textbf{0.8441} & {30.44/0.8433}   \\ 
& 50  & {27.50/0.7464} & 27.44/0.7458 & {27.41/0.7499} & 27.42/\textbf{0.7502} \\ 
& 75  & {25.89/0.6881} & 25.80/0.6880 & {25.80/0.6947} & 25.77/\textbf{0.6950} \\ \hline
& 1 & {23.06/0.5958} & 22.99/0.5949  & 23.00/\textbf{0.6050} & 22.95/0.6038 \\
Poisson & 2  & {24.23/0.6403} & 24.17/0.6377 & 24.15/\textbf{0.6501} & 24.15/0.6488  \\ 
& 4 &  {25.39/0.6858} & 25.33/0.6829 & 25.30/\textbf{0.6911} & 25.31/0.6899  \\
& 8 &  {26.77/0.7332} & 26.72/0.7329 & 26.71/\textbf{0.7388} & 26.71/0.7371 \\ \hline
& 0.1/1 & {22.73/0.5938} & 22.65/0.5936 & 22.64/\textbf{0.6044} & 22.60/0.6019 \\
& 0.2/2 & {23.75/0.6345} & 23.69/0.6337 & 23.68/\textbf{0.6402} & 23.66/0.6400 \\
& 0.5/5 & {25.10/0.6878} & 24.98/0.6860 & 24.95/\textbf{0.6955} & 24.91/0.6933 \\
Poisson-Gaussian & 1/10 & {26.14/0.7263} & 26.07/0.7255 & 26.05/\textbf{0.7334} & 26.05/0.7330 \\
& 2/20 &  {27.29/0.7613} & 27.18/0.7600 & 27.15/\textbf{0.7677} & 27.15/0.7659 \\
& 3/30 &  {27.77/0.7785} & 27.64/0.7770 & 27.59/\textbf{0.7844} & 27.61/0.7840 \\
& 6/60 & {28.59/0.8010} & 28.51/0.8001 & {28.50/0.8068} & 28.50/\textbf{0.8077} \\
& 12/120 & {28.88/0.8147} & 28.80/0.8122 & 28.77/\textbf{0.8180} & 28.78/0.8166 \\ \hline
\end{tabular}
\end{table*}

\section{Experiments}
\subsection{Experiment setting}
\noindent{}For image denoising, we use the PASCAL VOC 2010 dataset \cite{everingham2010pascal}. We follow the same training and testing split as \cite{remez2017deep}, 1,000 testing images are used to evaluate the performance of the proposed DN-ResNet, while the remaining images are used for training. Random Gaussian/Poisson/Poisson-Gaussian noisy images are generated with different noise levels. We consider AWGN with different noise variances $\sigma^2$, where $\sigma \in \{10, 25, 50, 75\}$. We follow the same way as \cite{azzari2016variance}\cite{remez2017deep2}, before corrupting with Poisson noise,  the input image pixel values are scaled to have  a max peak value from the set $peak \in\{1,2,4,8\}$. For the Poisson-Gaussian noise, we follow the same setting as \cite{makitalo2013optimal}, where $\sigma \in \{0.1, 0.2, 0.5, 1, 2, 3, 6, 12\}, peak=10\times{}\sigma$. $33\times{}33$ noisy patches and the corresponding $17\times{}17$ clean patches are cropped for training. For more comparison to other existing methods, we also use the Set14 \cite{zeyde2010single}, and BSD \cite{martin2001database} datasets in the testing.

For compressed image restoration, we utilize the dataset provided by the challenge on learned image compression (CLIC) \cite{CLIC}.  The commonly-used image compression algorithms, JPEG, JPEG 2000, and BPG (Better Portable Graphics) are utilized to obtain the decoded images. $33\times{}33$ decoded patches and the corresponding $17\times{}17$ clean patches are further extracted for training. 

Our networks are trained on y/cb/cr channels\footnote{In the quantitative evaluation, we show the PSNR/SSIM of the networks trained on y-channel only for fair comparison to existing work.}. For non-blind denoising, multiple networks are trained for each noise level respectively. In contrast, only one DN-ResNet is trained for blind denoising by mixing all training samples corrupted by Gaussian/Poisson/Poisson-Gaussian noises. Peak-Signal-to-Noise-Ratio (PSNR) and SSIM are utilized as evaluation protocol.

\begin{table*}[ht]
\caption{PSNR (dB)/SSIM evaluation of 13-layer DN-ResNets with different ResBlocks for blind denoising. `DN' is DN-ResNet, `DS-DN' is DS-DN-ResNet constructed by incremental evolution from DN-ResNet. `DS-DN-os' is the DS-DN-ResNet constructed by one-shot fine-tuning from DN-ResNet. `edge-a' denotes that the network is trained by edge-aware loss function. MACs are calculated for $640\times480$ input. }
\scriptsize
\renewcommand{\arraystretch}{1.0}
\centering
\begin{tabular}{|c|c|c|c|c|c|c|}
\hline
{DN-ResNet} & sigma/peak  & {DN} & {DS-DN} & DS-DN-os&DN+`edge-a' & DS-DN+`edge-a'\\ \hline
Parameters & - & 149,344 & 63,728 & 63,728 & 149,344 &  63,728 \\ 
MACs (Billion) & - & 45.9 & 19.6 & 19.6 & 45.9 & 19.6 \\ \hline
& 10 & 34.88/0.9217 & 34.79/0.9206 & 34.41/0.9088 & {34.88/0.9271} & 34.79/0.9259  \\
Gaussian  & 25 & 30.47/0.8369 & 30.36/0.8355 & 30.00/0.8240 & {30.45/0.8441} & {30.36/0.8433}   \\ 
& 50  & 27.44/0.7458 & 27.34/0.7439 & 26.99/0.7298 &  {27.41/0.7499} & 27.32/0.7484 \\ 
& 75  & 25.80/0.6880 & 25.74/0.6878 & 25.32/0.6759& {25.80/0.6947} & 25.72/0.6939 \\ \hline
& 1 & 22.99/0.5949  & 22.89/0.5933 & 22.59/0.5870 & {23.00/0.6050} & 22.89/0.6040 \\
Poisson & 2   & 24.17/0.6377 & 24.11/0.6364 & 23.77/0.6302 & {24.15/0.6501} & 24.09/0.6499  \\ 
& 4  & 25.33/0.6829 & 25.26/0.6811 & 24.88/0.6733 & {25.30/0.6911} & 25.24/0.6899  \\
& 8  & 26.72/0.7329 & 26.62/0.7314 & 26.30/0.7265 & {26.71/0.7388} & 26.60/0.7377 \\ \hline
& 0.1/1  & 22.65/0.5936 & 22.57/0.5919 & 22.17/0.5830 & {22.64/0.6044} & 22.56/0.6038 \\
& 0.2/2  & 23.69/0.6337 & 23.58/0.6322 & 23.20/0.6269 & {23.68/0.6402} & 23.54/0.6389 \\
& 0.5/5  & 24.98/0.6860 & 24.93/0.6843 & 24.65/0.6788 & {24.95/0.6955} & 24.93/0.6941 \\
Poisson- & 1/10 & 26.07/0.7255 & 25.99/0.7239 & 25.66/0.7188 & {26.05/0.7334} & 26.00/0.7330 \\
-Gaussian & 2/20  & 27.18/0.7600 & 27.12/0.7596 & 26.80/0.7524 & {27.15/0.7677} & 27.11/0.7666 \\
& 3/30  & 27.64/0.7770 & 27.57/0.7755 &  27.24/0.7700 & {27.59/0.7844} & 27.53/0.7839 \\
& 6/60 & 28.51/0.8001 & 28.46/0.7991 & 28.16/0.7929 & {28.50/0.8068} & 28.45/0.8061 \\
& 12/120  & 28.80/0.8122 & 28.74/0.8108 & 28.44/0.8059 & {28.77/0.8180} & 28.68/0.8177 \\ \hline
\end{tabular}
\end{table*}

\begin{figure}[h]
\centering
\includegraphics[height=5cm]{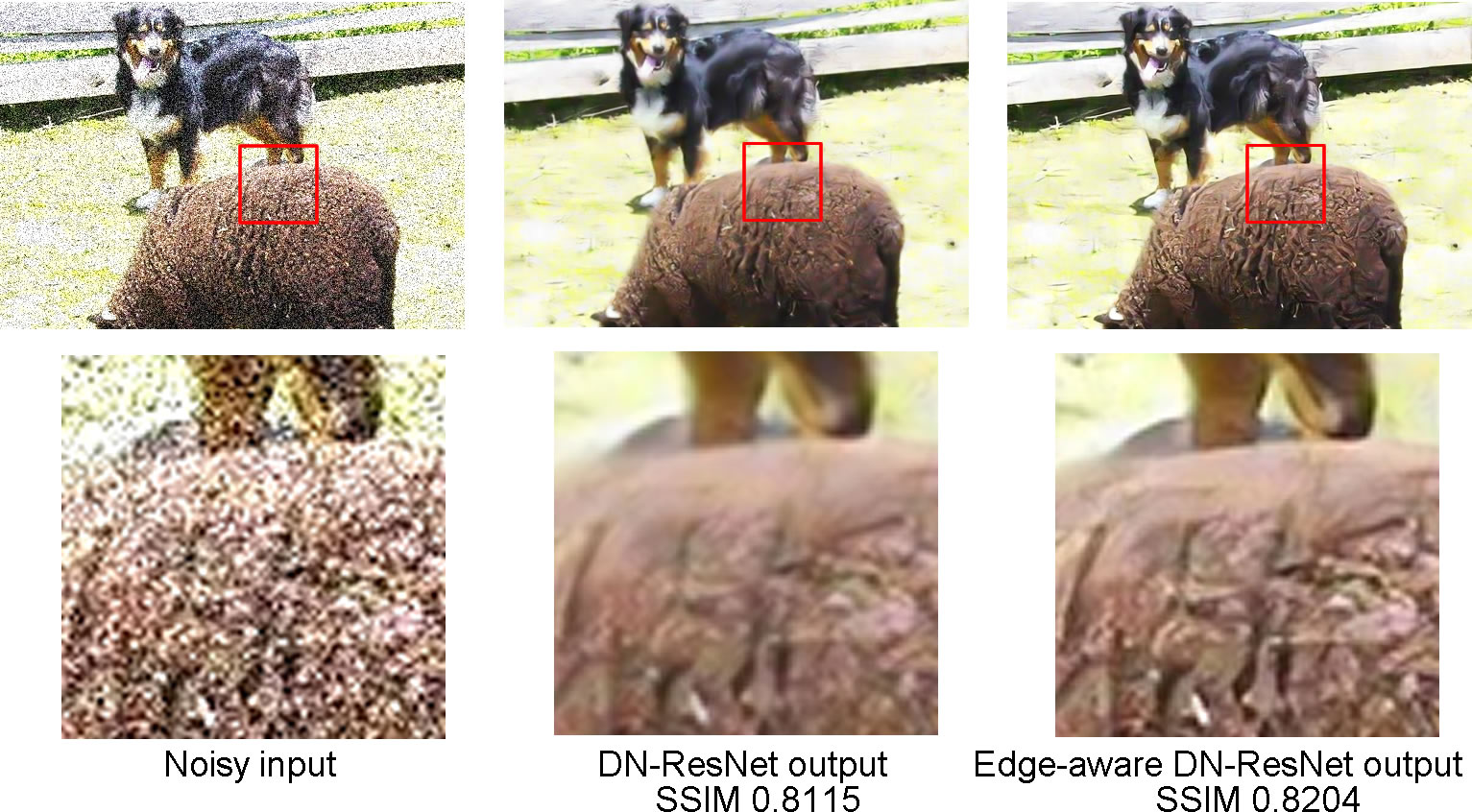}
\caption{Example outputs of different Poisson-Gaussian blind denoising networks. Left:noisy input. Mid:DN-ResNet output. Right:edge-aware DN-ResNet output. }
\label{fig:example}
\end{figure}

\subsection{Experiments on image denoising}
\noindent{}We first test the DN-ResNets up to 13 layers on non-blind Gaussian, Poisson, and Poisson-Gaussian denoising. These DN-ResNets are trained by cascading the ResBlocks in Fig. 1(b). The conventional MSE loss is utilized for all networks. In Table 1, we find that for all the above three denoising problems, the PSNR consistently increases along with using more layers. Although the deepest network we show is 13-layer DN-ResNet, the accuracy could still be further improved by cascading more layers. This is consistent with `the deeper, the better'. We also compare the cascade training versus one-shot training (`13-layer-os' in Table 1), where an end-to-end 13-layer DN-ResNet is trained from unsupervised weight initialization. We observe that such one-shot training results in 0.3 dB PSNR degradation compared to cascade training. This result makes sense because cascade training can be considered as a `partial-supervised weight initialization', its convergence will be easier compared to one-shot training based on unsupervised weight initialization.

Next, we test the DN-ResNets trained by edge-aware loss function described in Section 3.3, as well as utilizing DN-ResNet for blind denoising. In Table 2, we observe that utilizing DN-ResNet for blind denoising will not decrease the accuracy much compared to non-blind denoising. This trade-off is valuable since blind denoising does not require a time-consuming noise level estimation. In addition, we show that utilizing edge-aware loss function (blind+`edge-a'/`edge-b') improves the SSIM 0.005-0.01, without degrading the PSNR much. Since the conventional MSE has the same equation as PSNR, the slightly degradation in PSNR of the edge-aware DN-ResNet is reasonable. Using the edge map generated by Sobel gradient magnitude (blind+`edge-a', $w=0.025$ in Eq. (5)) is better than binary edge mask (blind+`edge-b', $w=4$). The perceptive quality is clearly improved as well, as illustrated in Fig. 6. It can be seen that the output from edge-aware DN-ResNet has sharper edge and higher SSIM compared to the output from ordinary DN-ResNet. This shows the effectiveness of emphasizing edge pixels during the training.\

Moreover, we evaluate the DN-ResNets constructed by different ResBlocks for the blind denoising problem. In Table 3, We observe that the DS-DN-ResNet (DS-DN) decreases less than 0.1 dB PSNR and less than 0.002 SSIM compared to DN-ResNet, but the computational cost (MACs, number of multiplications and accumulations) and the network size are significantly reduced. We also notice that if the DS-DN-ResNet is constructed by one-shot fine-tuning DN-ResNet (DS-DN-os), both the PSNR and SSIM will decrease a lot. This indicates that the proposed incrementally evolved DS-DN-ResNet is able to improve the efficiency of DN-ResNet. Using the DS-ResBlock together with the edge-aware loss function, we can achieve considerable accuracy, good perceptive quality, and less computational cost at the same time.

\begin{table}[h]
\caption{PSNR (dB)/SSIM comparison to the state of art Gaussian/Poisson/Poisson-Gaussian denoising algorithms on PASCAL VOC 2010 dataset. }
\scriptsize
\renewcommand{\arraystretch}{1.0}
\centering
\begin{tabular}{|c|c|c|c|c|c|}
\hline
{Gaussian sigma} &  blind & {10} & {25} & {50}  & {75} \\ \hline
BM3D \cite{dabov2007image} & No &  34.26/0.9197 & 29.62/0.8294 & 26.61/0.7404 & 25.12/0.6852 \\
DN-CNN-3 \cite{zhang2017beyond} & Yes  & - & 29.87/0.8350 & 26.85/0.7439 & - \\ 
DN-CNN-S \cite{zhang2017beyond}  & No & 34.79/0.9216 & 30.23/0.8379 & 27.29/0.7444 & 25.58/0.6888 \\ 
DenoiseNet \cite{remez2017deep} & No &  34.87/0.9219 & 30.36/0.8388 & 27.32/0.7447 & 25.74/0.6899 \\
DN-ResNet-13+`edge-a'  & Yes & \textbf{34.88/0.9271} & \textbf{30.45/0.8441} & \textbf{27.41/0.7499} & \textbf{25.80/0.6947} \\ 
DS-DN-ResNet-13+`edge-a' & Yes & 34.79/0.9259 & 30.36/0.8433 & {27.32/0.7484} & {25.72/0.6939} \\  \hline
{Poisson peak} & blind & {1} & {2} & {4}  & {8} \\ \hline
IVST+BM3D \cite{azzari2016variance} & No & 22.71/0.5920 & 23.70/0.6418 & 24.78/0.6815 & 26.08/0.7297 \\
DenoiseNet \cite{remez2017deep2} & No & 22.87/0.5989 & 24.09/0.6452 & 25.26/0.6857 & 26.70/0.7329 \\
DN-ResNet-13+`edge-a' & Yes & \textbf{23.00/0.6050} & \textbf{24.15/0.6501} & \textbf{25.30/0.6911} & \textbf{26.71/0.7388} \\
DS-DN-ResNet-13+`edge-a' & Yes & 22.89/0.6040 & 24.09/0.6499 & 25.24/0.6899 & 26.60/0.7377 \\  \hline
{Poisson-Gaussian sigma/peak} & blind & {0.1/1} & {0.2/2} & {0.5/5}  & {1/10}  \\ \hline
GAT+BM3D \cite{makitalo2013optimal} & No & 21.28/0.5451 & 22.56/0.5795 & 24.13/0.6478 & 25.38/0.7008  \\
DN-ResNet-13+`edge-a' & Yes & \textbf{22.64/0.6044} & \textbf{23.68/0.6402} & \textbf{24.95/0.6905} & \textbf{26.05/0.7334}  \\ 
DS-DN-ResNet-13+`edge-a' & Yes & {22.56/0.6038} & {23.54/0.6389} & {24.93/0.6941} & {26.00/0.7330}  \\ \hline
{Poisson-Gaussian sigma/peak} & blind &  2/20 & 3/30 & 6/60 & 12/120 \\ \hline
GAT+BM3D \cite{makitalo2013optimal} & No  & 26.50/0.7249 & 27.07/0.7587 & 27.87/0.7849 & 28.43/0.7962 \\
DN-ResNet-13+`edge-a' & Yes & \textbf{27.15/0.7677} & \textbf{27.59/0.7844} & \textbf{28.50/0.8068} & \textbf{28.77/0.8180} \\ 
DS-DN-ResNet-13+`edge-a' & Yes & 27.11/0.7666 & 27.53/0.7839 & 28.45/0.8061 & 28.68/0.8177 \\ \hline
\end{tabular}
\end{table}

\begin{figure*}[h]
\centering
\includegraphics[height=6.5cm]{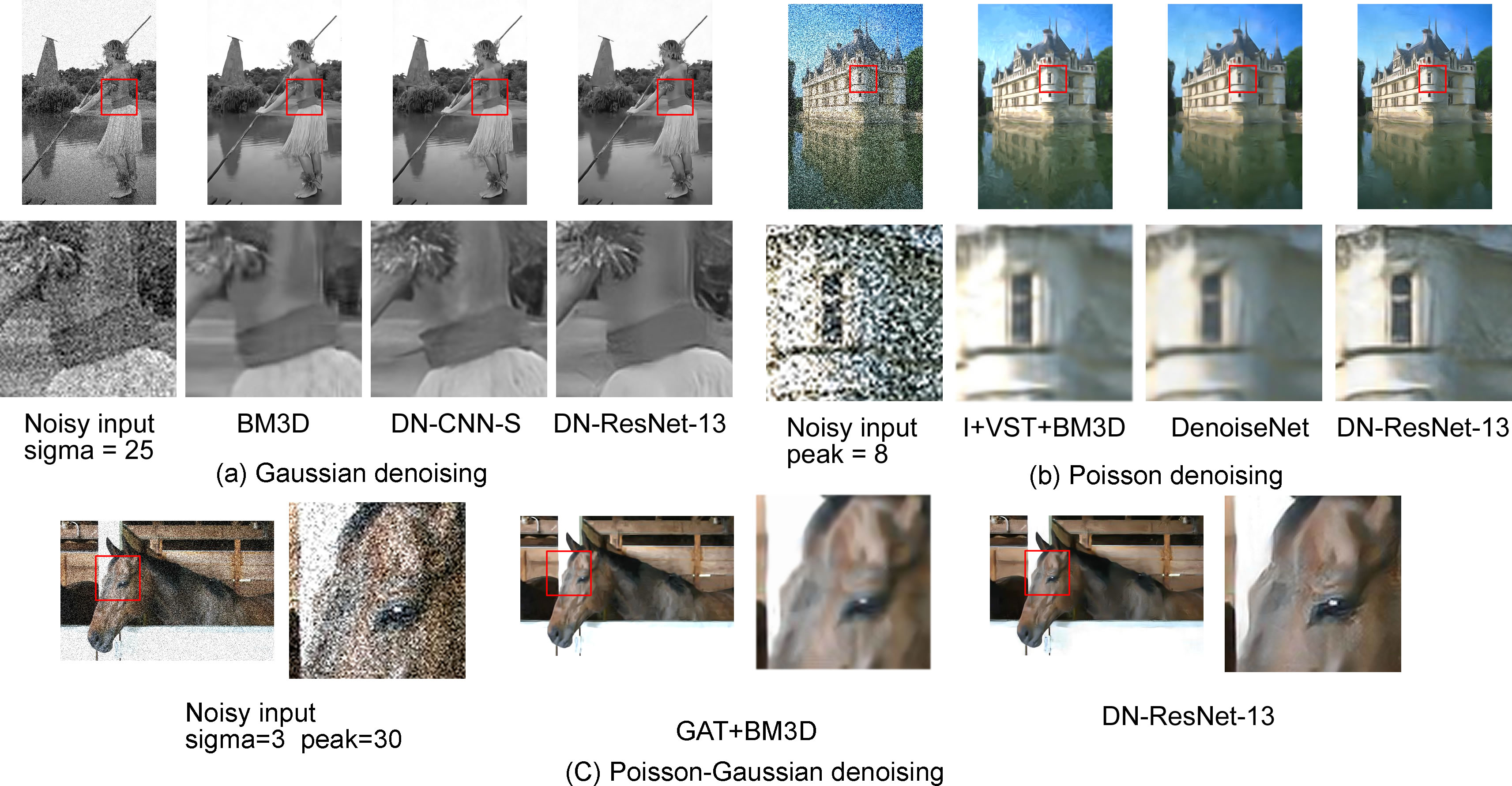}
\caption{Example outputs of different algorithms for Gaussian, Poisson, and Poisson-Gaussian denoising.}
\label{fig:example}
\end{figure*}

\begin{figure*}[h]
\centering
\includegraphics[height=6.5cm]{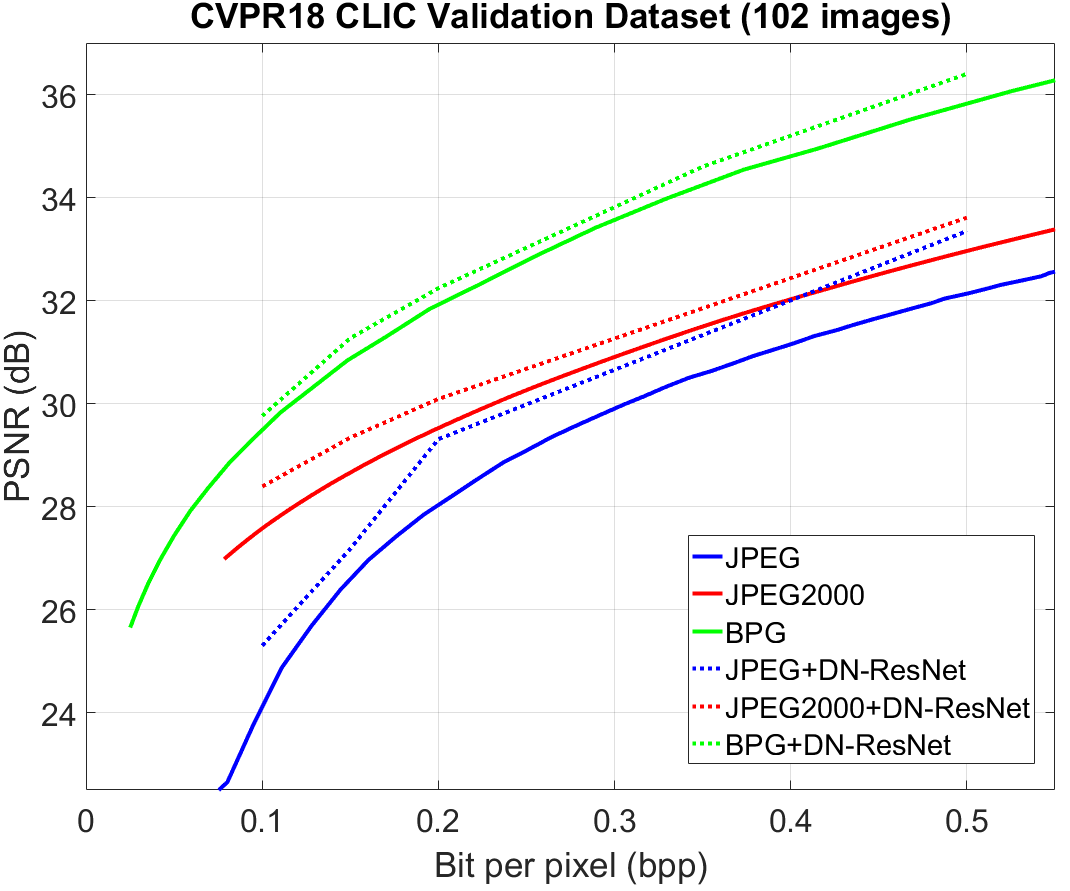}
\caption{Bit per pixel vs. PSNR in CLIC validation dataset.}
\label{fig:example}
\end{figure*}

\begin{table*}[h]
\caption{PSNR (dB)/SSIM evaluation of different blind Gaussian/Poisson denoising methods on multiple datasets. The speed is tested in single Titan-X GPU and $512\times512$ image.}
\scriptsize
\renewcommand{\arraystretch}{1.0}
\centering
\begin{tabular}{|c|c|c|c|c|c|}
\hline
{Gaussian sigma=50} & {Set14} & {BSD200} & VOC2010 &{Network Size} & Speed(ms) \\ \hline
MemNet \cite{tai2017memnet} &26.99/0.7794&25.89/0.7207&27.02/0.7422&667K & 343.28 \\
DN-CNN \cite{zhang2017beyond} &27.05/0.7788&25.83/0.7214&26.85/0.7439&650K & 55.44 \\
DN-ResNet-13+`edge-a' &\textbf{27.15/0.7849}&\textbf{25.99/0.7270}&\textbf{27.41/0.7499}&149K& 17.92\\
DS-DN-ResNet-13+`edge-a' &27.05/0.7826&25.87/0.7247&27.32/0.7484&\textbf{64K} & \textbf{8.33} \\ \hline
{Poisson peak=8} & {Set14} & {BSD200} & VOC2010 &{Network Size}   & Speed(ms)\\ \hline
MemNet \cite{tai2017memnet} &25.14/0.7198&25.77/0.7031&26.34/0.7321&667K& 343.28 \\
DN-ResNet-13+`edge-a'&\textbf{25.25/0.7242}&\textbf{25.91/0.7075}&\textbf{26.71/0.7388}&149K& 55.44\\
DS-DN-ResNet-13+`edge-a'&25.10/0.7220&25.79/0.7061&26.60/0.7374&\textbf{64K} & \textbf{8.33}\\ \hline
\end{tabular}
\end{table*}


\begin{figure*}[h]
\centering
\includegraphics[height=8cm]{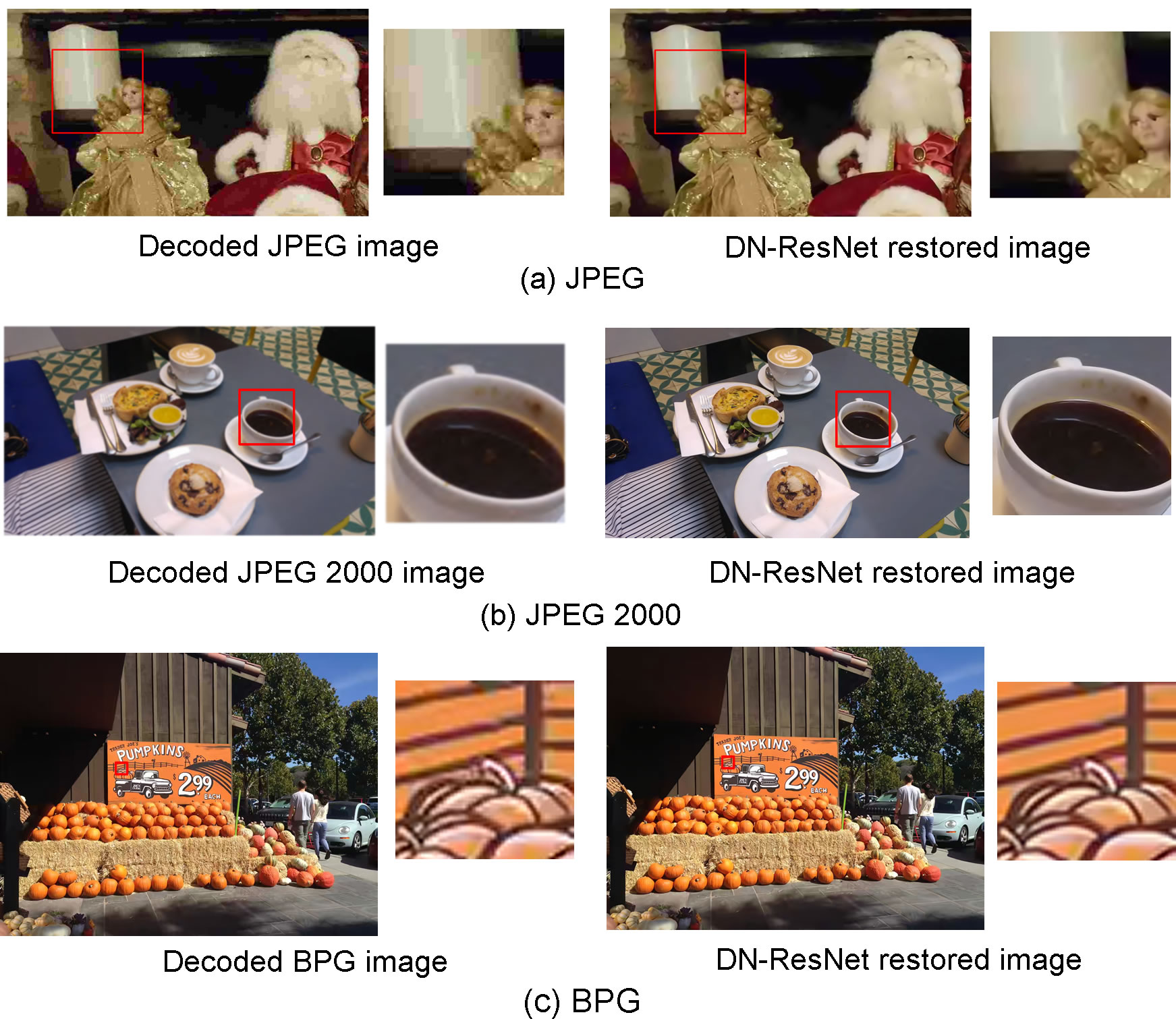}
\caption{Example outputs of DN-ResNet for compressed image restoration at 0.15 bpp.}
\label{fig:example}
\end{figure*}

\subsection{Comparison to the state of art denoising algorithms}
\noindent{}In Table 4, we compare the proposed DN-ResNet to the state of art denoising algorithms in PASCAL VOC dataset. For fair comparison, we retrain other networks using the same VOC dataset. We observe that DN-ResNet-13 blind denoising network clearly outperforms other blind and non-blind Gaussian denoising algorithms. Compared to the 20-layer DN-CNN-S \cite{zhang2017beyond}, DenoiseNet \cite{remez2017deep}, and MemNet  \cite{tai2017memnet} which contain more than 600K parameters, DN-ResNet achieves competitive performance, but the network size (150K parameters) is x4 times smaller. DN-ResNet takes 15-20ms to process a $512\times512$ image on single Titan X GPU, compared to 50-60ms for DN-CNN and DenoiseNet. DS-DN-ResNet only takes 8-10ms to process a $512\times512$ image, with the cost of less than 0.1 dB accuracy loss. These results show the effectiveness of DN-ResNet for Gaussian denoising. Example outputs are given in Fig. 7.


In Table 5, we also give the Gaussian and Poisson denoising results on other datasets, Set14, and BSD. The observation is consistent to PASCAL VOC datasets, where DN-ResNet and DS-DN-ResNet still achieve better accuracy. As summary, the proposed DN-ResNet and DS-DN-ResNet achieve the state of art performance for Gaussian/Poisson/Poisson-Gaussian denoising, with better efficiency and smaller network size compared to existing deep CNNs. They are effective for both blind denoising and non-blind denoising.

\subsection{Experiments on compressed image restoration}
\noindent{}Besides image denoising, we also evaluate the proposed DN-ResNet on compressed image restoration. In Fig. 8, the curves of compression ratio (bpp, bit per pixel) versus PSNR of the decoded compressed image and restored image are given. We can find that DN-ResNet is able to improve the quality of the decoded images for all compression methods. 1-2 dB, 0.5-0.7 dB, and 0.3-0.4 dB gain can be observed for JPEG, JPEG 2000, and BPG respectively. Fig. 9 shows some restored images at 0.15 bit per pixel, where DN-ResNet clearly improves the perceptive quality of the decoded compressed images.

Acutally, our network can also be applied for image super resolution. We cascade our DN-ResNet to 19 layers and apply it for image super resolution \cite{ren2017image}. The low-resolution images are considered as noisy input, and the high-resolution images are considered as clean image. We observe that our DN-ResNet also achieves better PSNR and SSIM for all the scale 2,3,4 in Set 5 and Set 14, but the network size is still 1/3 compared to existing networks such as MemNet \cite{tai2017memnet} or DRRN \cite{tai2017image}. This indiates that our DN-ResNet can be utilized as an unified framework for image enhancement.



\section{Conclusion}
\noindent{}In this paper, we presented the DN-ResNet for image denoising achieving both high accuracy and efficiency. We show that cascade training is effective in training efficient deep ResNets. The perceptive quality can be enhanced by using edge-aware loss function. We further propose the depthwise separable ResBlock and incrementally evolve the DN-ResNet to DS-DN-ResNet. The computational cost of DN-ResNet is reduced 2.5 times, with less than 0.1 dB PSNR loss. Experimental results on benchmark datasets show that for either blind or non-blind denoising, the proposed DN-ResNet achieves better accuracy and efficiency compared to the state of art denoising networks on all types of noises, including Gaussian, Poisson, and Poisson-Gaussian. The same network architecture can be utilized for other image enhancement applications as well.


\end{document}